\documentclass{moriond}

\usepackage{amsmath}

\def\be{\begin{equation}}
\def\ee{\end{equation}}
\def\bea{\begin{eqnarray}}
\def\eea{\end{eqnarray}}

\newcommand{\Photo}{}

\begin{document}
\vspace*{4cm}
\title{Catching UHE Neutrinos with HERON}

\author{A. Zeolla on behalf of the HERON Collaboration}

\address{Sorbonne Université et CNRS, UMR 7095, Institut d'Astrophysique de Paris,\\ 98 bis bd Arago, 75014 Paris, France}

\maketitle\abstracts{
The Hybrid Elevated Radio Observatory for Neutrinos, or HERON, is a newly proposed ultrahigh energy Earth-skimming tau neutrino detector. Ultrahigh energy tau neutrinos which skim the Earth may produce $\tau$-leptons which escape into the atmosphere and initiate up-going extensive air showers. The HERON concept consists of 24 compact phased radio arrays, embedded within a larger sparse array of 360 standalone antennas, distributed along a mountain range and designed to capture the radio emission of these up-going extensive air showers. Due to the high elevation observation sites and the long propagation length of radio, HERON achieves a very large instantaneous effective area towards the horizon, and thus excels at the detection of astrophysical transient events such as gamma-ray bursts. With the excellent pointing resolution offered by the sparse array, HERON would be capable of conducting UHE neutrino astronomy and could be incorporated into the broader network of multi-messenger instruments. Here, we detail the HERON concept and describe the science which can be accomplished with it.}

\section{Motivation}

Ultrahigh energy ($E>100$ PeV) neutrinos are of great interest because they could potentially identify the sources of ultrahigh energy cosmic rays, as well as probe fundamental particle physics at energies unattainable with man-made accelerators \cite{Ackermann:2019ows,Ackermann:2019cxh}. The IceCube Neutrino Observatory and the Pierre Auger Observatory, however, have both constrained the ultrahigh energy diffuse neutrino flux to very low levels \cite{IceCubeCollaborationSS:2025jbi,AbdulHalim:2023SN}. Only one ultrahigh energy neutrino candidate has been detected to-date, a 120 PeV muon neutrino detected in 2023 by KM3NeT \cite{KM3NeT:2025npi}. The source of this neutrino is currently unknown.

Many very large-scale neutrino detectors (IceCube Gen-2, GRAND, etc.) are currently envisioned which can achieve the sensitivity necessary to probe the tightly-constrained UHE diffuse flux, however, the construction of these detectors will take at least a decade. In the meantime, it seems advantageous to target astrophysical fluxes if we wish to detect UHE neutrinos, particularly those of astrophysical transients. Many astrophysical neutrino flux models (active galactic nuclei, binary neutrino star mergers, gamma-ray bursts, etc.) peak near a neutrino energy of 100 PeV \cite{Kotera:2025jca}. An experiment designed to measure these fluxes should therefore maximize its sensitivity in this region. This is additionally advantageous since it serves to bridge the gap between IceCube and the future very large-scale detectors. Lastly, for the greatest chance of discovering UHE neutrinos from a transient event, the detector should have a very large instantaneous effective area.

One strategy for detecting UHE neutrinos is the radio detection of up-going extensive air showers (EAS) generated by Earth-skimming tau neutrinos. At ultrahigh energies, tau neutrinos which skim the Earth can create a $\tau$-lepton which then exits the Earth before quickly decaying in the atmosphere. This decay initiates an up-going EAS, which in turn emits radio due to the geomagnetic effect. Radio has an in-air propagation length of $\mathcal{O}$(100 km), leading to an exciting prospect: a single radio array could monitor a very large area for the radio impulses created by these Earth-skimming events. By combining the observations of multiple arrays, one can thus achieve sensitivity to the UHE flux in an efficient manner. Experiments which seek to detect UHE neutrinos using this technique include BEACON \cite{BEACON} and GRAND \cite{GRAND}.

\section{The HERON Experiment}

\begin{figure}[tbp]
  \centering
  \includegraphics[width=0.8\textwidth]{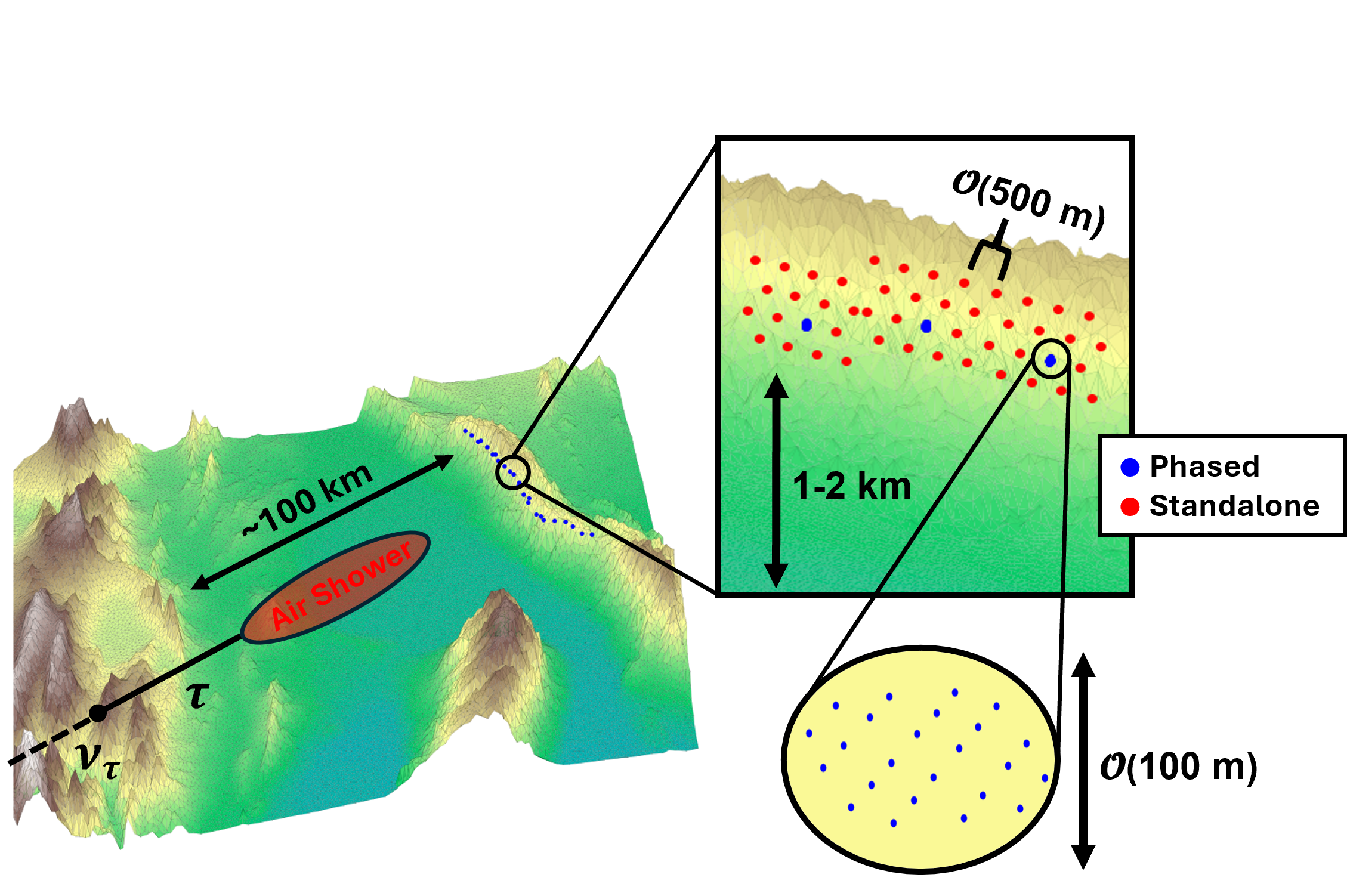}
  \caption{The baseline HERON concept: an Earth-skimming tau neutrino produces an up-going EAS within a valley. 24 phased arrays, each containing 24 antennas, are arranged along a mountain range, monitoring the valley for the radio impulse generated by the EAS. The phased arrays are embedded within a larger sparse array consisting of 360 standalone antennas.}
  \label{fig:heron}
\end{figure}

The Hybrid Elevated Radio Observatory for Neutrinos, or HERON, combines aspects of the BEACON and GRAND experiments to target the flux of Earth-skimming tau neutrinos at 100 PeV. The baseline design of HERON consists of 24 phased radio arrays, each containing 24 antennas, embedded within a larger sparse array of 360 standalone antennas. The phased arrays are distributed along the side of a mountain range with a spacing of $2-3$ km and altitudes of $1-2$ km. This spacing reduces the overlap in individual detection areas when viewing a neutrino source, maximizing the effective area. The increased number of antennas per phased array and the lower altitude relative to BEACON ($3-4$ km) were chosen to improve the sensitivity at 100 PeV. Each phased array is relatively compact, with a maximum antenna separation of $\mathcal{O}(100 \, \text{m})$, while the standalone antennas are spaced $\mathcal{O}(500 \, \text{m})$ apart. This ensures that during the detection of an up-going EAS, the antennas within a phased array will receive nearly identical signals, while those of the sparse array will image different regions of the radio footprint. A schematic of the HERON concept is shown in Fig.~\ref{fig:heron}.

Digital beamforming, in which signals are delayed according to programmed time delays (beams) and then summed, allows the phased array to trigger on signals that otherwise would not be recorded. Even signals which are individually sub-dominant to noise may produce a prominent impulse when summed across 24 channels. In this way, phasing lowers the energy threshold and enhances the sensitivity of the instrument. Each highly-sensitive phased array passes its triggers onto the 15 nearest standalone antennas, collecting their signals for offline event reconstruction. Individual beam thresholds are automatically adjusted such that a $\sim10$ Hz global trigger rate is maintained per phased array. In the event of a multi-messenger alert, the beam pattern can be redefined in real time such that a beam points in the direction of a target-of-opportunity and the beam threshold can be lowered to enhance the chance of discovery.

The standalone antennas image the radio footprint. Distinguishing features, such as the Cherenkov ring, serve to validate the detection of an extensive air shower. Offline interferometric reconstruction enables the sparse array to trace out the particle axis. Initial simulations suggests that HERON could achieve a $<0.4^\circ$ pointing resolution with this method \cite{Zeolla:2025anb}, enabling UHE neutrino astronomy. Interferometric reconstruction can also be used to determine the location of $X_\text{max}$, essential for determining the parent particle and its energy. The radio emission of an EAS is highly beamed in the forward direction, thus each event should only be viewed by a subset of the antennas. Signals which are present across the entire sparse array are likely anthropogenic backgrounds. In this way, the standalone antennas can also provide a veto for EAS detection. Lastly, we plan to implement an independent standalone trigger which could be used to detect cosmic rays, for example.

The topography seen in Fig.~\ref{fig:heron} is the actual topography present at the proposed experiment site in San Juan province, Argentina. Shown on the left side of the figure is a portion of the Sierra del Tontal mountain range, while the antennas are shown distributed along the Sierra de Valle F\'ertil mountain range on the right. We are still investigating which site is better suited for deployment. A mountain range running roughly north-south is ideal since the geomagnetic effect is maximized to the east and west. Between the two mountain ranges is a smooth valley about 100 km wide. The valley is mostly undeveloped, except for a few very small towns. Radio background measurements performed at the site are promising.

\section{Projected Performance}

\begin{figure}
\centering
\includegraphics[width=0.50\textwidth,trim={1.3cm 0cm 3.5cm 0cm},clip]{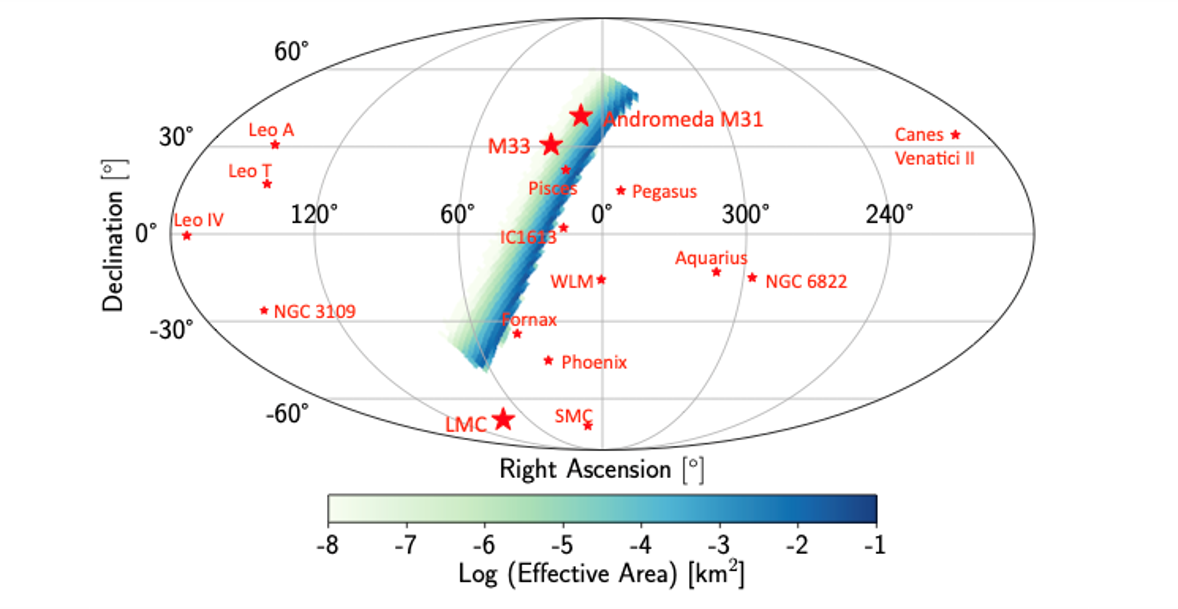}
\includegraphics[width=0.49\textwidth,]{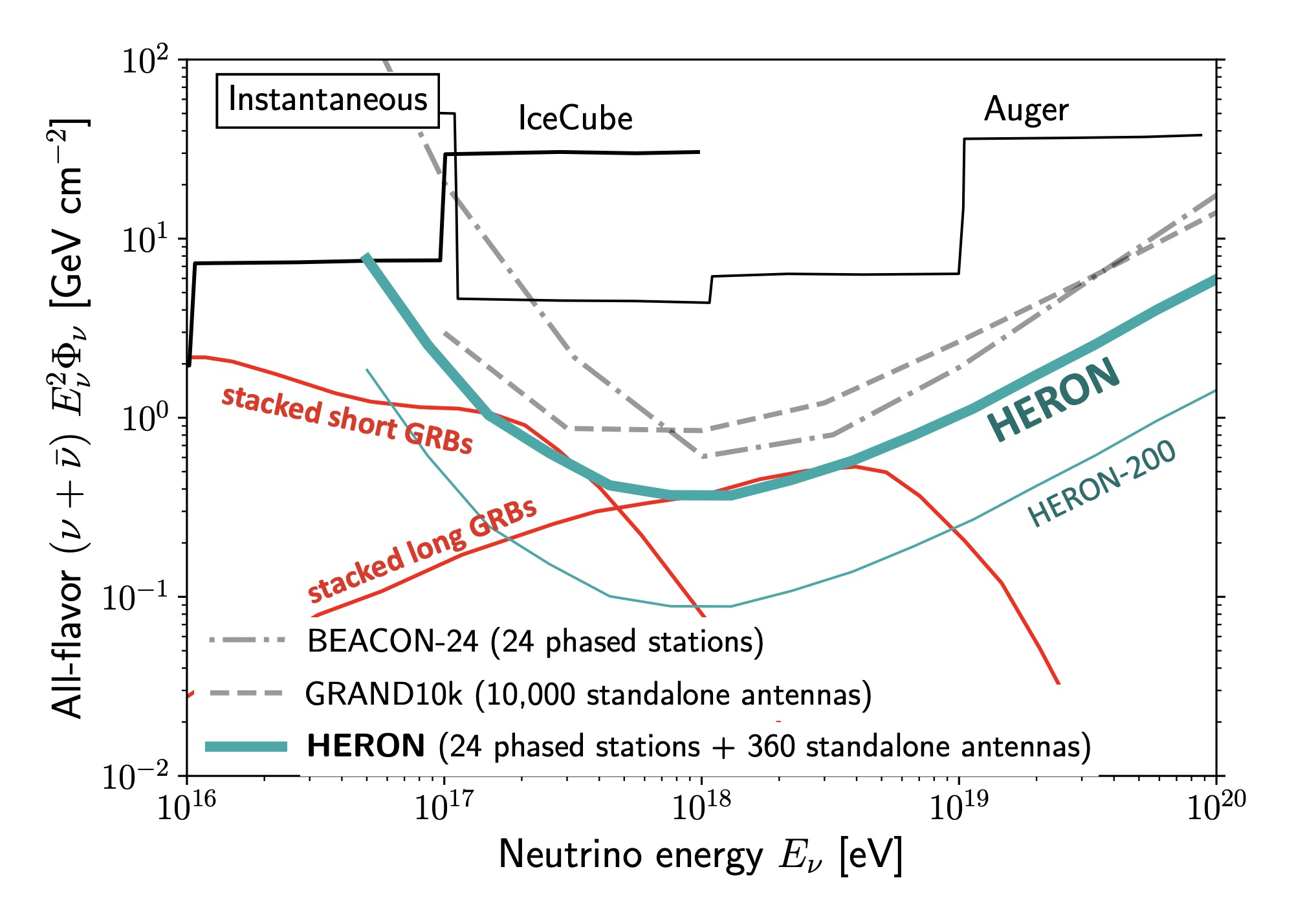}
    \caption{\textbf{Left:} Instantaneous effective area of HERON at 100 PeV. \textbf{Right:} All-flavor neutrino sensitivity for short-duration ($<15$ min) astrophysical transient events \protect\cite{GRAND:2025rps}.}
    \label{fig:short}
\end{figure}

The simulated instantaneous effective area of HERON at 100 PeV is shown on the left in Fig.~\ref{fig:short}, as a function of right-ascension and declination. For an Earth-skimming tau neutrino detector, the instantaneous effective area is maximized just below the horizon. This is because the $\tau$-lepton exit probability is maximized at these cord lengths, while a huge area is simultaneously in view of the arrays. As the elevation angle of a source sinks below the horizon, both exit probability and the geometric area in-view simultaneously decrease, and the instantaneous effective area rapidly drops. We see that for an elevated radio observatory like HERON, we thus end up with a very large instantaneous effective area in a narrow region of the sky. On the right of Fig.~\ref{fig:short}, we convert the maximum instantaneous effective area as a function of neutrino energy into a short-burst neutrino sensitivity. We see that HERON improves on existing limits set by IceCube and the Pierre Auger Observatory by a factor of 10. Additionally, we see that HERON has sufficient sensitivity to potentially detect gamma-ray bursts.

\begin{figure}
\centering
\includegraphics[width=0.50\textwidth,trim={1.3cm 0cm 4cm 0cm},clip]{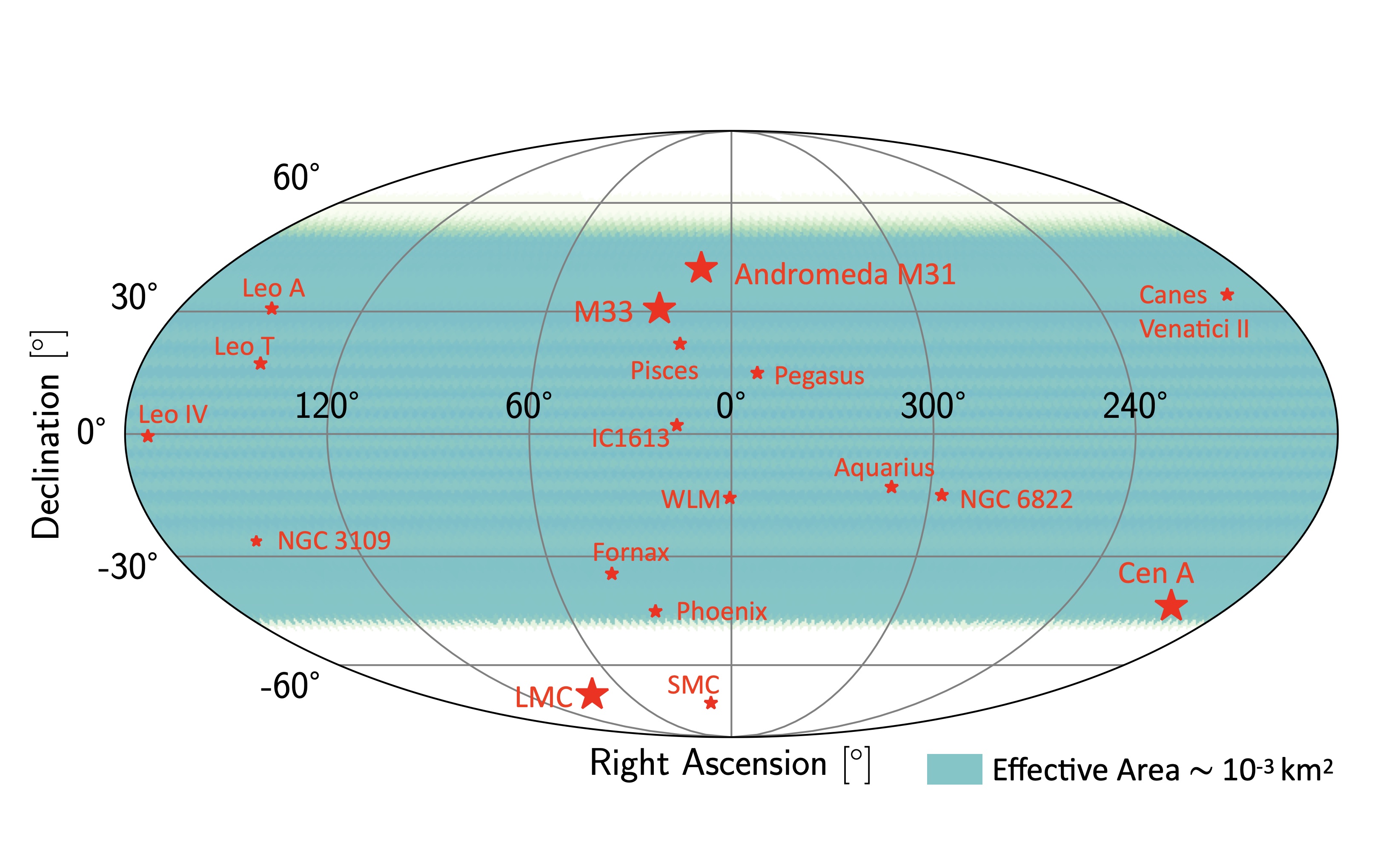}
\includegraphics[width=0.49\textwidth,]{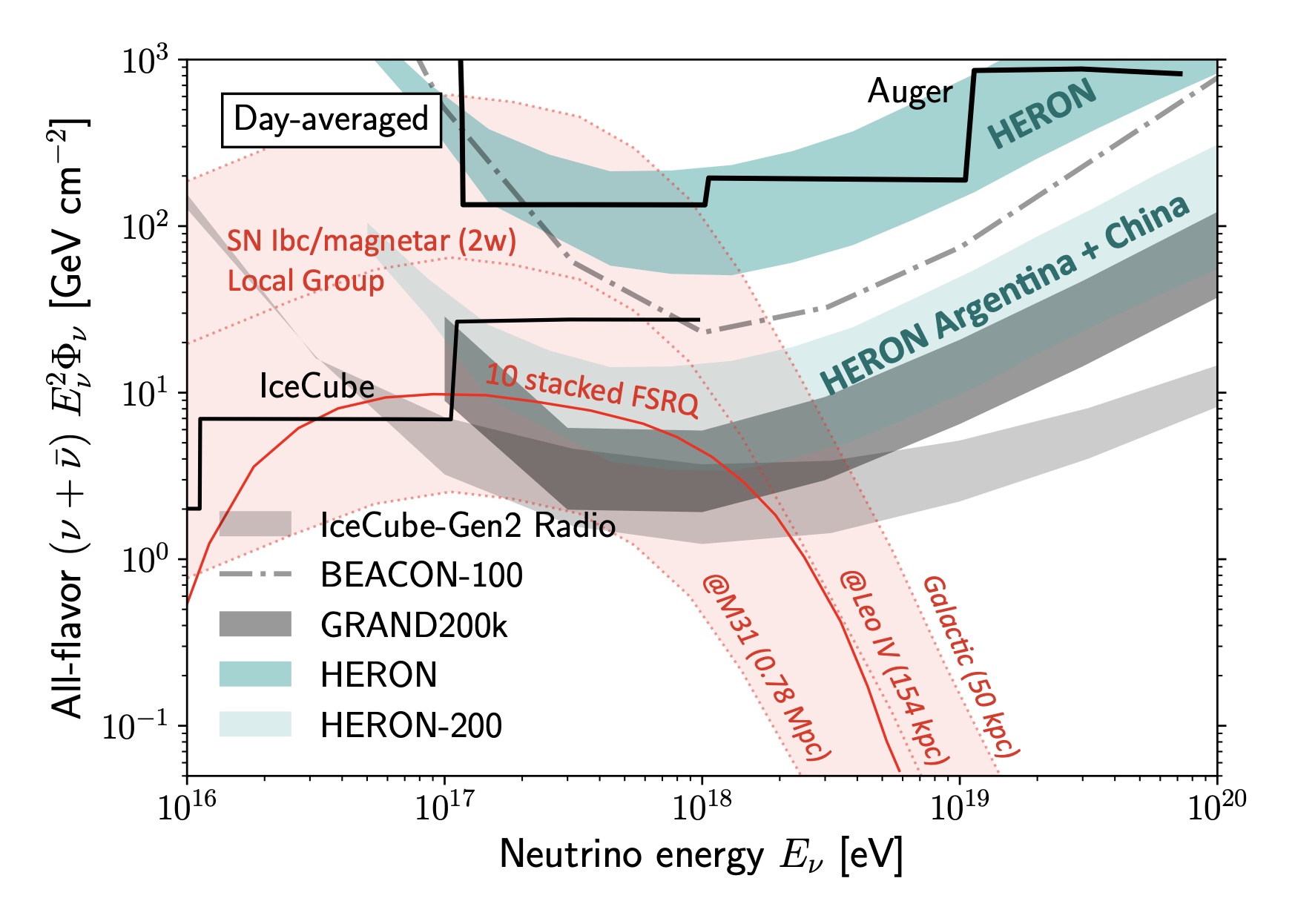}
    \caption{\textbf{Left:} Day-average effective area of HERON at 100 PeV. \textbf{Right:} All-flavor neutrino sensitivity for long-duration ($>1$ day) astrophysical transient events \protect\cite{GRAND:2025rps}.}
    \label{fig:long}
\end{figure}

As the Earth rotates, the instantaneous effective area will sweep across the sky in right-ascension. The resulting day-average effective area is shown on the left in Fig.~\ref{fig:long}. We see that because HERON faces east (or west) from a position relatively near the equator, about $\sim70\%$ of the sky is observed each day. With such large sky-coverage, HERON is likely to be able to follow-up on many multi-messenger alerts. On the right of Fig.~\ref{fig:long}, the day-average effective area is converted into a long-burst neutrino sensitivity. We see that HERON matches existing limits set by Auger and can potentially detect local newly-born magnetars. A hypothetical 200-array HERON would improve upon existing limits set by IceCube and could conduct stacked-searches for neutrinos from FSRQs.

The large effective area achieved by placing antenna arrays at high elevation provides HERON with excellent sensitivity to transient neutrino fluxes. With just 936 total antennas, HERON achieves sensitivity at ultrahigh energies in a highly efficient manner. At the time of its completion, HERON will have the best short-burst neutrino sensitivity above 100 PeV. This, combined with its pointing resolution and reconstruction capabilities, will enable HERON to conduct UHE neutrino astronomy.

\section*{Acknowledgments}

This work was supported by the CNRS Programme Blanc MITI (``GRAND'' 2023.1 268448), the
CNRS Programme AMORCE ("GRAND" 258540; France), and the HERON project, Agence Nationale de la Recherche (ANR-24-MRS2-0014). K.\,K. acknowledges support from the Fulbright-France program. HERON has been awarded a Synergy Grant by the ERC.



\section*{References}
\bibliography{moriond}

@article{Ackermann:2019ows,
    author = "Ackermann, Markus and others",
    journal = "Bull. Am. Astron. Soc.",
    volume = "51",
    pages = "185",
    year = "2019"
}

@article{Ackermann:2019cxh,
    author = "Ackermann, Markus and others",
    journal = "Bull. Am. Astron. Soc.",
    volume = "51",
    pages = "215",
    year = "2019"
}

@article{IceCubeCollaborationSS:2025jbi,
    author = "Abbasi, R. and others",
    collaboration = "(IceCube Collaboration){\textsection}, IceCube",
    journal = "Phys. Rev. Lett.",
    volume = "135",
    number = "3",
    pages = "031001",
    year = "2025"
}

@article{AbdulHalim:2023SN,
  author = "Abdul Halim and others",
  journal = "PoS",
  year = "2023",
  volume = "ICRC2023",
  pages = "1488"
}

@article{KM3NeT:2025npi,
    author = "Aiello, S. and others",
    collaboration = "KM3NeT",
    journal = "Nature",
    volume = "638",
    number = "8050",
    pages = "376--382",
    year = "2025",
    note = "[Erratum: Nature 640, E3 (2025)]"
}

@article{Kotera:2025jca,
    author = "Kotera, Kumiko and others",
    journal = "JCAP",
    volume = "01",
    pages = "027",
    year = "2026"
}

@article{BEACON,
    author = "Wissel, Stephanie and others",
    journal = "JCAP",
    volume = "11",
    pages = "065",
    year = "2020"
}

@article{GRAND,
    author = "{\'A}lvarez-Mu{\~n}iz, Jaime and others",
    collaboration = "GRAND",
    journal = "Sci. China Phys. Mech. Astron.",
    volume = "63",
    number = "1",
    pages = "219501",
    year = "2020"
}

@article{GRAND:2025rps,
    author = "Kotera, Kumiko and others",
    journal = "PoS",
    volume = "ICRC2025",
    pages = "1078",
    year = "2025"
}

@article{Zeolla:2025anb,
    author = "Zeolla, Andrew and others",
    journal = "PoS",
    volume = "ICRC2025",
    pages = "1227",
    year = "2025"
}


\end{document}